\documentclass[11pt,preprint]{aastex}

\usepackage{longtable}

\newcommand{\etal}{{\it et~al.}}

\begin{document}

\title{The Increasingly Strange Polarimetric Behavior of the Barbarian Asteroids}

\author{Joseph R. Masiero\altaffilmark{1}, Maxime Devog\`{e}le\altaffilmark{2}, Isabella Macias\altaffilmark{3}, Joahan Castaneda Jaimes\altaffilmark{4},  Alberto Cellino\altaffilmark{5}}

\altaffiltext{1}{Caltech/IPAC, 1200 E California Blvd, MC 100-22, Pasadena, CA 91125, USA, {\it jmasiero@ipac.caltech.edu}}
\altaffiltext{2}{University of Central Florida}
\altaffiltext{3}{University of Florida}
\altaffiltext{4}{Caltech}
\altaffiltext{5}{INAF, Osservatorio Astrofisico di Torino, Italy}

\begin{abstract}

  Polarization phase-curve measurements provide a unique constraint on
  the surface properties of asteroids that are complementary to those
  from photometry and spectroscopy, and have led to the identification
  of the ``Barbarian'' asteroids as a class of objects with highly
  unusual surfaces.  We present new near-infrared polarimetric
  observations of six Barbarian asteroids obtained with the WIRC+Pol
  instrument on the Palomar Hale telescope.  We find a dramatic change
  in polarimetric behavior from visible to near-infrared for these
  objects, including a change in the polarimetric inversion angle that
  is tied to the index of refraction of the surface material.  Our
  observations support a two-phase surface composition consisting of
  high albedo, high index of refraction inclusions with a small
  optical size scale embedded in a dark matrix material more closely
  related to C-complex asteroids.  These results are consistent with
  the interpretation that the Barbarians are remnants of a population
  of primitive bodies that formed shortly after CAIs.  Near-infrared
  polarimetry provides a direct test of the constituent grains of
  asteroid surfaces.

\end{abstract}

\section{Introduction}

Polarimetric observations offer us a unique way of probing the
physical properties of the regolith of asteroids in our Solar system.
These observations complement results obtained via spectroscopy and
photometric modeling, providing an independent constraint on the
albedo, particle size, and index of refraction of the surface
materials.  Asteroid polarimetric properties are generally closely
linked with the taxonomic classification from spectral measurements,
allowing for a common interpretation of the measured properties.
Recent reviews of asteroid polarimetry by \citet{belskaya15},
\citet{cellino15}, and \citet{bagnulo17} highlight the current state
of knowledge and the outstanding questions in the field.

The measured percent polarization $P_r$ for an asteroid is a function
of the observational phase angle $\alpha$ (the angle between the Sun
and the telescope as seen by the asteroid).  The percent polarization
is defined with respect to the Sun-asteroid-Earth scattering plane,
with positive values being assigned to polarization perpendicular to
the plane and negative values to polarization in the scattering plane.
Scattered light follows a general trend of being unpolarized at
$\alpha=0^\circ$, reaching negative values between $-1\%$ and $-2\%$
(depending upon the surface albedo) at phase angles below
$\alpha<20^\circ$, recovering to $P_r=0\%$ at the inversion angle
$\alpha_{inv}\sim20^\circ$ and increasing to a maximum polarization at
$\alpha\sim90^\circ$.  This maximum polarization value, however, is
not observable for asteroids that are distant from the Sun, such as
those in the Main Belt, due to the restrictions of orbital geometry;
Main Belt asteroids typically can only be observed out to phases of
$\alpha\sim30^\circ$.

One of the most intriguing results to come out of surveys of asteroid
polarization at visible wavelengths was the discovery of objects with
very large inversion angles.  The first was identified by
\citet{cellino06}, with other members of the Ld-class spectroscopic
taxonomy soon after being found to have similar behavior
\citep[e.g.][etc]{gilhutton08,masiero09a,gilhutton11}.  As only a
subset of the Ld near-infrared spectroscopic class shows this unusual
behavior, these objects are referred to as Barbarians after the
first-discovered and namesake object (234) Barbara.
\citet{devogele18} present a detailed analysis of the combined
polarimetric and spectroscopic properties of these objects and show
evidence for a connection between the Barbarians and an enrichment of
spinel-bearing inclusions. They also found for (234) Barbara that the
inversion angle is dependent on the wavelength of observation, with
inversion angle ranging from 26$^\circ$ to 30$^\circ$ from the I (0.79
$\mu$m) to the B (0.44 $\mu$m) band respectively. They interpret this
change as a result of the refractive index of spinel which varies from
$n=$1.81 to $n=$1.78 in those bands, respectively \citep[cf.][]{hosseini08}.

The unusual polarimetric properties of the Barbarians, with the
implication of a surface rich in spinels, suggests that the
composition of these objects is dominated by some of the most
early-forming material in the Solar system.  One possible
interpretation of this is that the Barbarians are relics of the
earliest stages of protoplanet growth.  Here, we probe the composition
of the Barbarians using newly obtained near-infrared polarimetry in
comparison with visible-light polarimetry to test possible surface
compositions for these unusual objects.

\section{Observations and Data Reduction}

To conduct our study, we obtained polarization phase curves for six
different Barbarian asteroids at $J$ and $H$ bands ($1.25~\mu$m and
$1.64~\mu$m, respectively) using the Wide-field InfraRed
Camera+Polarimeter (WIRC+Pol) on the Palomar 200-inch telescope
\citep{tinyanont19a}.  WIRC+Pol employs a polarizing grating to split
the incoming light into its four linear polarization components, as
well as dispersing it into a low-resolution spectrum.  In addition,
WIRC+Pol has a half-wave plate (HWP) that can be rotated up-stream of the
other optics to swap the beams of each polarization state.  This
beam-swapping allows for a dramatic reduction is systematics, enabling
polarimetric accuracies of order $0.1\%$ \citep{tinyanont19b}.

Our survey follows the same procedure as described in
\citet{masiero22} and builds upon the initial results presented there.
In short, measurements are obtained at each of four HWP rotation
positions, and with the source at two different positions on the chip
to allow for A-B background subtraction.  Exposure times are scaled to
source brightness, with the longest individual exposure time of
30-seconds limited by the background flux.  For most objects we obtain
4 HWP cycles at each chip position for each filter, and this exposure
sequence takes 20-45 mins depending on the exposure time.  For the
faintest sources we obtain 6 or 8 HWP cycles, to improve our
measurement uncertainties.

Data reduction is carried out using the standard WIRC+Pol data
reduction pipeline
software\footnote{\textit{https://github.com/WIRC-Pol/wirc$\_$drp}}.
Dark images were obtained for each exposure time at the start and end
of each night, and flats were taken for both filters at the start and
end as well for calibration purposes.  The data reduction pipeline
produces Q/I and U/I polarization measurements across the observed
spectrum of the object, along with the total polarization (P) spectrum
and the polarization angle ($\theta$) spectrum.  To obtain the
single-band polarization values (and improve measurement uncertainty
for faint sources) we perform an error-weighted co-add of the Q/I and
U/I spectra and derive the total P and $\theta$ for each object, along
with associated uncertainties, shown in Table~\ref{tab.astdata}.  This
process of co-addition typically suppresses the statistical
uncertainty on polarization to values well below the $\sim0.1\%$
accuracy that external validation has demonstrated is achieved by the
system. To account for this we add $0.1\%$ in quadrature with the
derived statistical uncertainty to give us the final uncertainty
quoted in the table.  Analysis of the polarization spectral slope
across each band \citep[as was done by][]{kwon22} will be subject of
future investigation, but no obvious significant trends have been seen
to date for our asteroid observations.

\begin{figure}[ht]
\begin{center}
\includegraphics[scale=0.5]{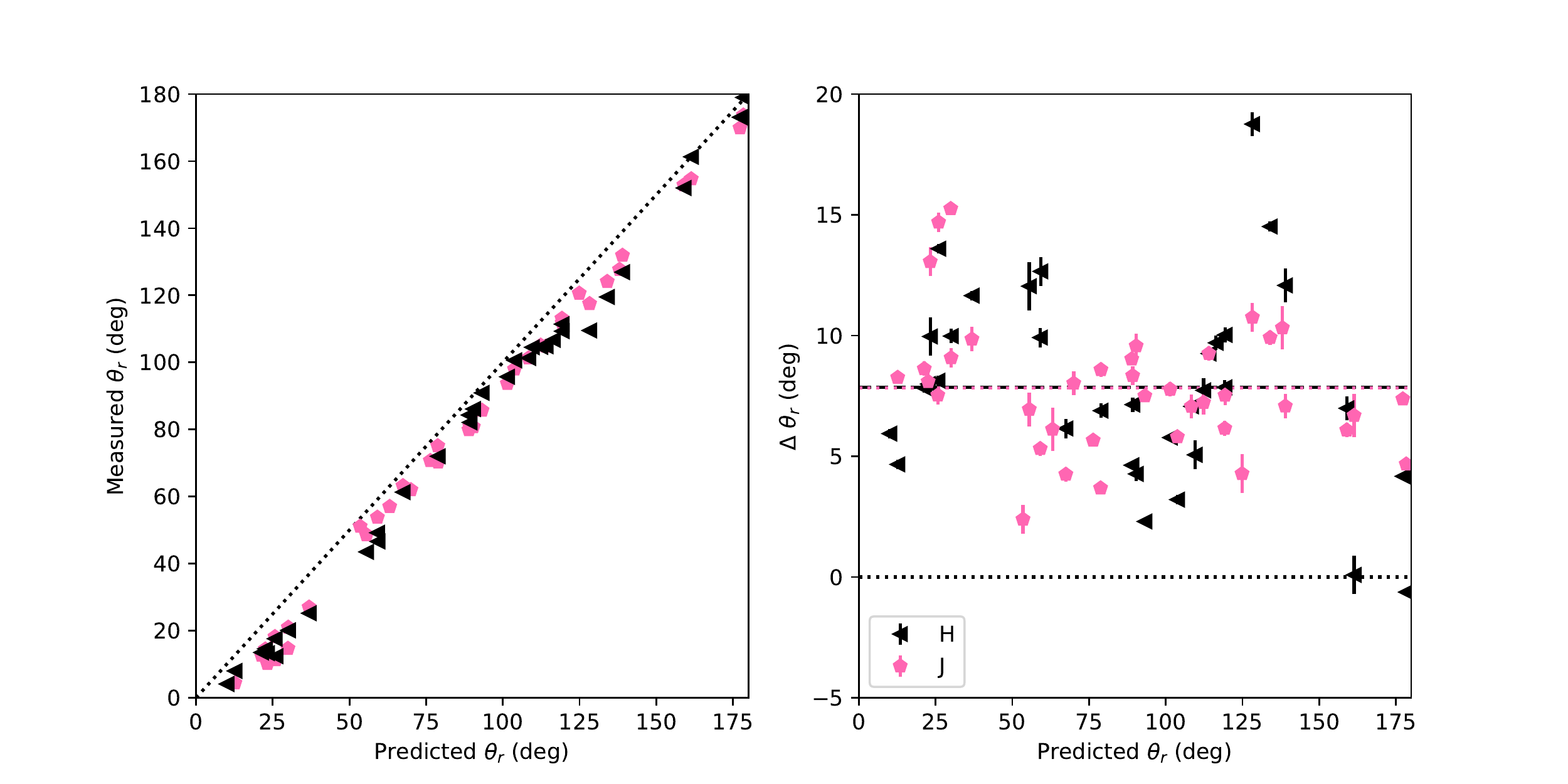}
\protect\caption{Left: Measured WIRC+Pol scattering angle $\theta_r$
  compared to the predicted scattering angle based on orbital geometry
  for all asteroids with polarization measurements $|P|>1$ for the J
  (pink pentagons) and H (black triangles) bands.  Right: Deviation in
  scattering angle (predicted - measured) compared to the predicted
  scattering angle.  The mean of all values in each band is shown by
  the dashed lines of the same color as the points, which overlap at
  $7.8^\circ$ for both bands.}
\label{fig.anglecal}
\end{center}
\end{figure}

We then rotate the angle of polarization such that $\theta_r=0$ is
defined as perpendicular to the scattering plane, resulting in a $P_r$
value that is positive when perpendicular to the plane and negative
when parallel to the plane, as is convention in asteroid polarimetry.
As discussed in \citet{millarblanchaer21} the calculated angle of
polarization $\theta$ is slightly offset from the expected value for
both $J$ and $H$ bands.  Following \citet{masiero22}, we take all our
measured values where $|P_r|>1\%$ and compare the expected angle of
polarization to the value reported by the data reduction pipeline.  We
choose these high polarization epochs to ensure that low level
systematic issues don't affect this correction.  We show the results
of this comparison in Figure~\ref{fig.anglecal}, and find a mean
offset of $\theta_{\rm expected}-\theta_{\rm measured} = 7.8^\circ$
for both the $J$ and $H$ bands; we apply this correction to obtain the
$\theta_{corr}$ values presented in Table~\ref{tab.astdata}.

\begin{table}[ht]
\begin{center}
\scriptsize
\noindent
{\tiny
\begin{longtable}{cccrrcccr}
\caption{WIRC+Pol Barbarian Results}\label{tab.astdata}\\
\hline\hline

Asteroid & Observation Date & UT & Total t$_{exp}$ (sec) & Phase (deg) & Filter & $P_r$ & $\theta_{corr}$ (E-of-N) & $\theta_{scattering plane}$\\ 

\hline\hline
\endfirsthead
\caption[]{(continued)}\\
\hline\hline
Asteroid & Observation Date & UT & Total t$_{exp}$ (sec) & Phase (deg) & Filter & $P_r$ & $\theta_{corr}$ (E-of-N) & $\theta_{scattering plane}$\\ 

\hline\hline
\endhead
\hline
\endfoot

234       & 2022-04-02 & 07:57 & 1920 &  9.3 & H & $-1.51 \pm 0.10\%$ & $134.7 \pm   2.0^\circ$ & $136.9^\circ$\\ 
234       & 2022-04-02 & 07:57 & 1920 &  9.3 & J & $-1.62 \pm 0.10\%$ & $139.7 \pm   1.8^\circ$ & $136.9^\circ$\\ 
234       & 2022-05-30 & 05:42 & 1920 & 20.8 & H & $-0.70 \pm 0.13\%$ & $107.9 \pm   4.4^\circ$ & $114.7^\circ$\\ 
234       & 2022-05-30 & 05:42 & 1920 & 20.8 & J & $-0.53 \pm 0.11\%$ & $123.2 \pm   6.0^\circ$ & $114.7^\circ$\\ 
387       & 2021-02-03 & 07:49 & 4800 &  5.7 & H & $-1.51 \pm 0.10\%$ & $ 92.2 \pm   1.9^\circ$ & $ 85.5^\circ$\\ 
387       & 2021-02-03 & 07:49 & 3000 &  5.7 & J & $-1.57 \pm 0.10\%$ & $ 87.6 \pm   1.9^\circ$ & $ 85.5^\circ$\\ 
387       & 2021-05-30 & 04:48 & 2640 & 19.7 & H & $-0.89 \pm 0.11\%$ & $109.3 \pm   3.3^\circ$ & $108.2^\circ$\\ 
387       & 2021-05-30 & 04:48 & 2640 & 19.7 & J & $-1.01 \pm 0.10\%$ & $110.4 \pm   2.9^\circ$ & $108.2^\circ$\\ 
387       & 2022-04-02 & 11:38 & 160  & 26.5 & H & $+0.43 \pm 0.10\%$ & $  2.8 \pm   6.8^\circ$ & $ 85.5^\circ$\\ 
387       & 2022-04-02 & 11:38 & 160  & 26.5 & J & $+0.27 \pm 0.11\%$ & $  5.5 \pm  10.8^\circ$ & $ 85.5^\circ$\\ 
387       & 2022-05-30 & 10:11 & 160  & 16.5 & H & $-0.97 \pm 0.10\%$ & $ 50.9 \pm   3.8^\circ$ & $ 52.6^\circ$\\ 
387       & 2022-05-30 & 10:11 & 160  & 16.5 & J & $-1.26 \pm 0.10\%$ & $ 59.0 \pm   2.4^\circ$ & $ 52.6^\circ$\\ 
387       & 2022-07-11 & 07:51 & 320  & 11.8 & H & $-1.67 \pm 0.10\%$ & $127.4 \pm   1.7^\circ$ & $129.9^\circ$\\ 
387       & 2022-07-11 & 07:51 & 320  & 11.8 & J & $-1.69 \pm 0.10\%$ & $132.0 \pm   1.7^\circ$ & $129.9^\circ$\\ 
387       & 2022-10-15 & 02:10 & 320  & 27.5 & H & $+0.45 \pm 0.11\%$ & $169.2 \pm   6.6^\circ$ & $ 83.8^\circ$\\ 
387       & 2022-10-15 & 02:10 & 320  & 27.5 & J & $+0.51 \pm 0.11\%$ & $179.3 \pm   5.8^\circ$ & $ 83.8^\circ$\\ 
980       & 2021-09-04 & 09:07 & 2880 & 17.1 & H & $-1.11 \pm 0.10\%$ & $ 27.9 \pm   2.6^\circ$ & $ 27.7^\circ$\\ 
980       & 2021-09-04 & 09:07 & 2880 & 17.1 & J & $-1.15 \pm 0.10\%$ & $ 28.6 \pm   2.6^\circ$ & $ 27.7^\circ$\\ 
980       & 2021-11-08 & 04:53 & 960  & 19.7 & H & $-0.61 \pm 0.10\%$ & $ 94.3 \pm   4.8^\circ$ & $ 87.9^\circ$\\ 
980       & 2021-11-08 & 04:53 & 960  & 19.7 & J & $-0.98 \pm 0.10\%$ & $ 85.7 \pm   3.0^\circ$ & $ 87.9^\circ$\\ 
980       & 2022-10-15 & 11:15 & 1920 & 19.6 & H & $-0.80 \pm 0.13\%$ & $104.5 \pm   3.7^\circ$ & $104.8^\circ$\\ 
980       & 2022-10-15 & 11:15 & 1440 & 19.6 & J & $-0.83 \pm 0.11\%$ & $107.8 \pm   3.6^\circ$ & $104.8^\circ$\\ 
980       & 2022-12-01 & 11:31 & 1920 & 17.8 & H & $-1.22 \pm 0.11\%$ & $109.0 \pm   2.4^\circ$ & $108.6^\circ$\\ 
980       & 2022-12-01 & 11:31 & 1920 & 17.8 & J & $-1.08 \pm 0.10\%$ & $109.1 \pm   2.7^\circ$ & $108.6^\circ$\\ 
980       & 2023-01-07 & 08:51 & 960  &  8.2 & H & $-1.61 \pm 0.10\%$ & $117.4 \pm   1.8^\circ$ & $120.9^\circ$\\ 
980       & 2023-01-07 & 08:51 & 960  &  8.2 & J & $-1.38 \pm 0.10\%$ & $119.3 \pm   2.1^\circ$ & $120.9^\circ$\\ 
236       & 2021-05-30 & 09:25 & 1920 & 19.7 & H & $-0.75 \pm 0.10\%$ & $ 71.8 \pm   4.0^\circ$ & $ 69.5^\circ$\\ 
236       & 2021-05-30 & 09:25 & 1920 & 19.7 & J & $-1.16 \pm 0.10\%$ & $ 69.8 \pm   2.6^\circ$ & $ 69.5^\circ$\\ 
236       & 2021-06-26 & 09:06 & 2880 & 12.7 & H & $-1.54 \pm 0.10\%$ & $ 56.7 \pm   1.9^\circ$ & $ 57.8^\circ$\\ 
236       & 2021-06-26 & 09:06 & 2880 & 12.7 & J & $-1.77 \pm 0.10\%$ & $ 61.5 \pm   1.7^\circ$ & $ 57.8^\circ$\\ 
236       & 2021-09-04 & 06:19 & 960  & 18.3 & H & $-1.17 \pm 0.11\%$ & $ 89.4 \pm   2.5^\circ$ & $ 88.6^\circ$\\ 
236       & 2021-09-04 & 06:19 & 960  & 18.3 & J & $-1.23 \pm 0.10\%$ & $ 88.6 \pm   2.4^\circ$ & $ 88.6^\circ$\\ 
236       & 2021-11-08 & 02:24 & 1440 & 25.1 & H & $+0.33 \pm 0.11\%$ & $158.7 \pm  10.1^\circ$ & $ 75.0^\circ$\\ 
236       & 2021-11-08 & 02:24 & 1440 & 25.1 & J & $+0.05 \pm 0.10\%$ & $141.1 \pm  57.4^\circ$ & $ 75.0^\circ$\\ 
236       & 2022-10-15 & 09:40 & 1920 & 21.9 & H & $-0.59 \pm 0.11\%$ & $ 99.3 \pm   5.0^\circ$ & $ 94.6^\circ$\\ 
236       & 2022-10-15 & 09:40 & 1920 & 21.9 & J & $-0.79 \pm 0.10\%$ & $ 92.6 \pm   3.8^\circ$ & $ 94.6^\circ$\\ 
236       & 2022-12-01 & 09:32 & 960  &  9.3 & H & $-2.20 \pm 0.10\%$ & $113.3 \pm   1.3^\circ$ & $116.3^\circ$\\ 
236       & 2022-12-01 & 09:32 & 960  &  9.3 & J & $-1.79 \pm 0.10\%$ & $113.1 \pm   1.7^\circ$ & $116.3^\circ$\\ 
236       & 2023-01-07 & 07:44 & 360  &  8.9 & H & $-1.57 \pm 0.10\%$ & $ 54.3 \pm   1.9^\circ$ & $ 61.9^\circ$\\ 
236       & 2023-01-14 & 05:48 & 960  & 11.2 & H & $-1.54 \pm 0.10\%$ & $ 69.0 \pm   1.9^\circ$ & $ 69.0^\circ$\\ 
236       & 2023-01-14 & 05:48 & 960  & 11.2 & J & $-2.04 \pm 0.10\%$ & $ 71.1 \pm   1.5^\circ$ & $ 69.0^\circ$\\ 
172       & 2022-07-11 & 11:05 & 1920 & 25.2 & H & $+0.11 \pm 0.10\%$ & $155.8 \pm  28.2^\circ$ & $ 74.9^\circ$\\ 
172       & 2022-07-11 & 11:05 & 1920 & 25.2 & J & $-0.23 \pm 0.10\%$ & $ 74.5 \pm  12.9^\circ$ & $ 74.9^\circ$\\ 
172       & 2022-10-15 & 07:10 & 1920 & 18.1 & H & $-0.86 \pm 0.10\%$ & $ 59.6 \pm   3.5^\circ$ & $ 62.5^\circ$\\ 
172       & 2022-10-15 & 07:10 & 1920 & 18.1 & J & $-1.07 \pm 0.10\%$ & $ 64.8 \pm   2.8^\circ$ & $ 62.5^\circ$\\ 
172       & 2022-12-01 & 08:33 & 160  &  9.0 & H & $-1.83 \pm 0.11\%$ & $117.1 \pm   1.7^\circ$ & $122.9^\circ$\\ 
172       & 2022-12-01 & 08:33 & 160  &  9.0 & J & $-1.62 \pm 0.10\%$ & $125.4 \pm   1.9^\circ$ & $122.9^\circ$\\ 
172       & 2023-01-07 & 06:17 & 960  & 20.1 & H & $-0.85 \pm 0.11\%$ & $ 90.6 \pm   3.4^\circ$ & $ 82.2^\circ$\\ 
172       & 2023-01-07 & 06:17 & 960  & 20.1 & J & $-0.94 \pm 0.10\%$ & $ 87.6 \pm   3.1^\circ$ & $ 82.2^\circ$\\ 
172       & 2023-01-14 & 04:01 & 960  & 21.5 & H & $-0.60 \pm 0.10\%$ & $ 75.7 \pm   5.0^\circ$ & $ 80.0^\circ$\\ 
172       & 2023-01-14 & 04:01 & 960  & 21.5 & J & $-1.02 \pm 0.11\%$ & $ 79.5 \pm   2.9^\circ$ & $ 80.0^\circ$\\ 
402       & 2022-04-02 & 08:53 & 160  &  8.8 & H & $-1.87 \pm 0.10\%$ & $ 51.3 \pm   1.8^\circ$ & $ 50.2^\circ$\\ 
402       & 2022-04-02 & 08:53 & 160  &  8.8 & J & $-1.78 \pm 0.10\%$ & $ 56.4 \pm   1.8^\circ$ & $ 50.2^\circ$\\ 
402       & 2022-05-30 & 07:05 & 1440 & 20.4 & H & $-1.00 \pm 0.11\%$ & $125.1 \pm   3.1^\circ$ & $124.5^\circ$\\ 
402       & 2022-05-30 & 07:05 & 1440 & 20.4 & J & $-1.22 \pm 0.10\%$ & $128.4 \pm   2.5^\circ$ & $124.5^\circ$\\ 
402       & 2022-07-11 & 05:18 & 1920 & 24.2 & H & $-0.27 \pm 0.14\%$ & $ 77.8 \pm  13.3^\circ$ & $112.3^\circ$\\ 
402       & 2022-07-11 & 05:18 & 1920 & 24.2 & J & $-0.43 \pm 0.11\%$ & $125.0 \pm   7.9^\circ$ & $112.3^\circ$\\

\hline
\end{longtable}
Polarization measurement $P_r$ has been rotated such that positive
values represent polarization perpendicular to the
Sun-Asteroid-Telescope scattering plane ($\theta_{scattering plane}$)
and negative values represent polarization in this
plane. $\theta_{corr}$ is the angle of polarization after
correcting for the observed $7.8^\circ$ systematic offset.
}
\end{center}
\end{table}

\clearpage

\section{Results}

\begin{figure}[ht]
\begin{center}
\includegraphics[scale=0.5]{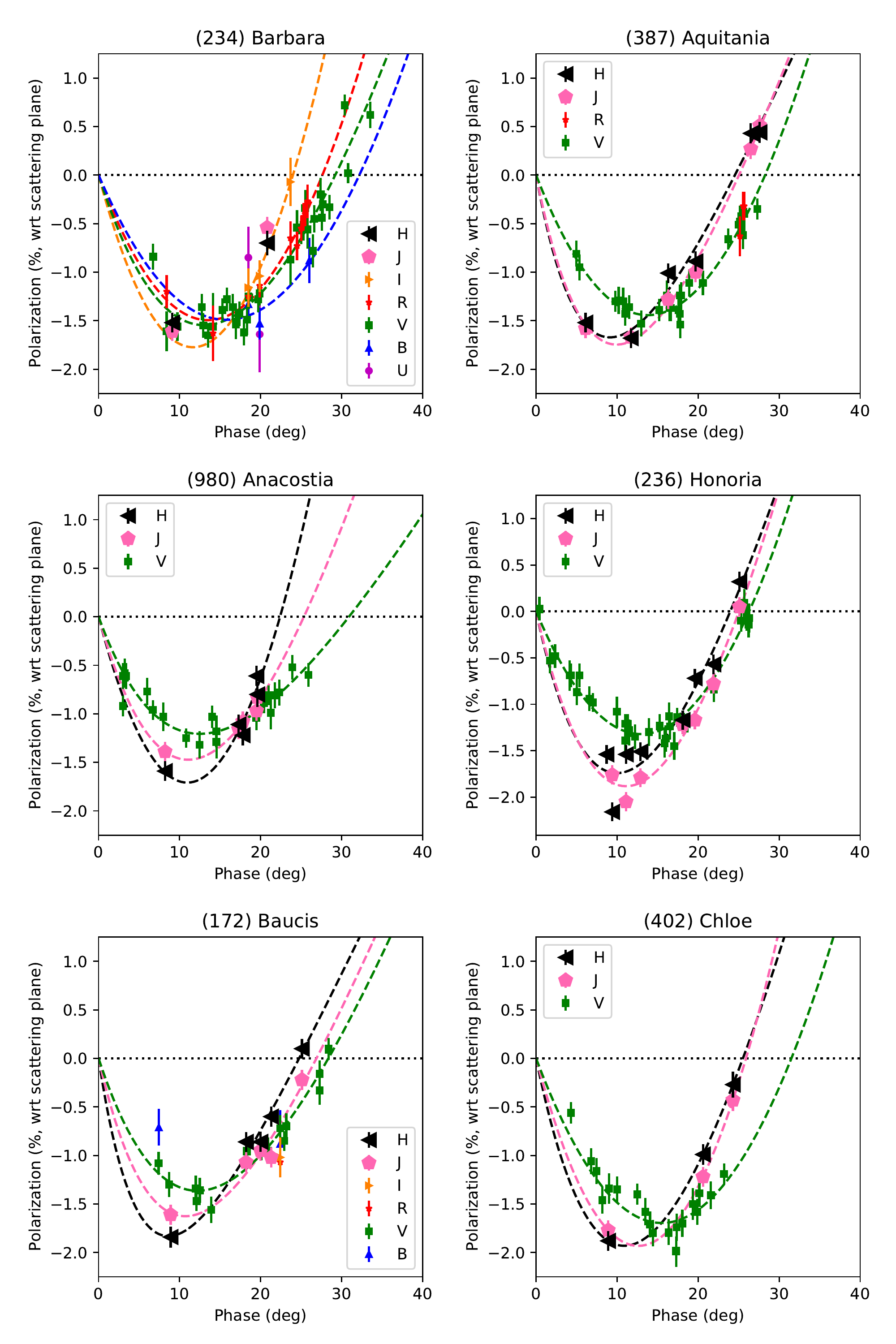}
\protect\caption{Polarization phase curves for the six Barbarian
  asteroids we observed in the near-infrared with WIRC+Pol ($J$ and
  $H$ bands).  Other polarization measurements for wavelengths plotted
  are drawn from the literature \citep{apd,bendjoya22}.  Phase
  function fits for any bandpass that had more than 3 observations are
  shown as dashed lines with colors matching those used to identify
  the data points.}
\label{fig.split}
\end{center}
\end{figure}

We show in Figure~\ref{fig.split} the polarization phase curves for
the six Barbarian asteroids that we have observed with WIRC+Pol.  We
also include all literature measurements at other wavelengths that are
available in three archives: the Asteroid Polarimetric Database
\citep{apd}, the Calern Asteroid Polarimetric Survey (CAPS), and the
CASLEO Survey \citep{bendjoya22}.  Following \citet{masiero22}, we fit
the analytical polarization phase function defined in
\citet{muinonen09} to each wavelength independently if that dataset
had at least three data points that spanned at least $5^\circ$ of
phase.  These are shown in the figure as dashed lines, with colors to
indicate wavelength following the color of the points in the legend.
All six objects show a consistent change from the $V$ band
observations (which dominate the literature measurements) to the $J$
and $H$ band data presented here.

\begin{figure}[ht]
\begin{center}
\includegraphics[scale=0.6]{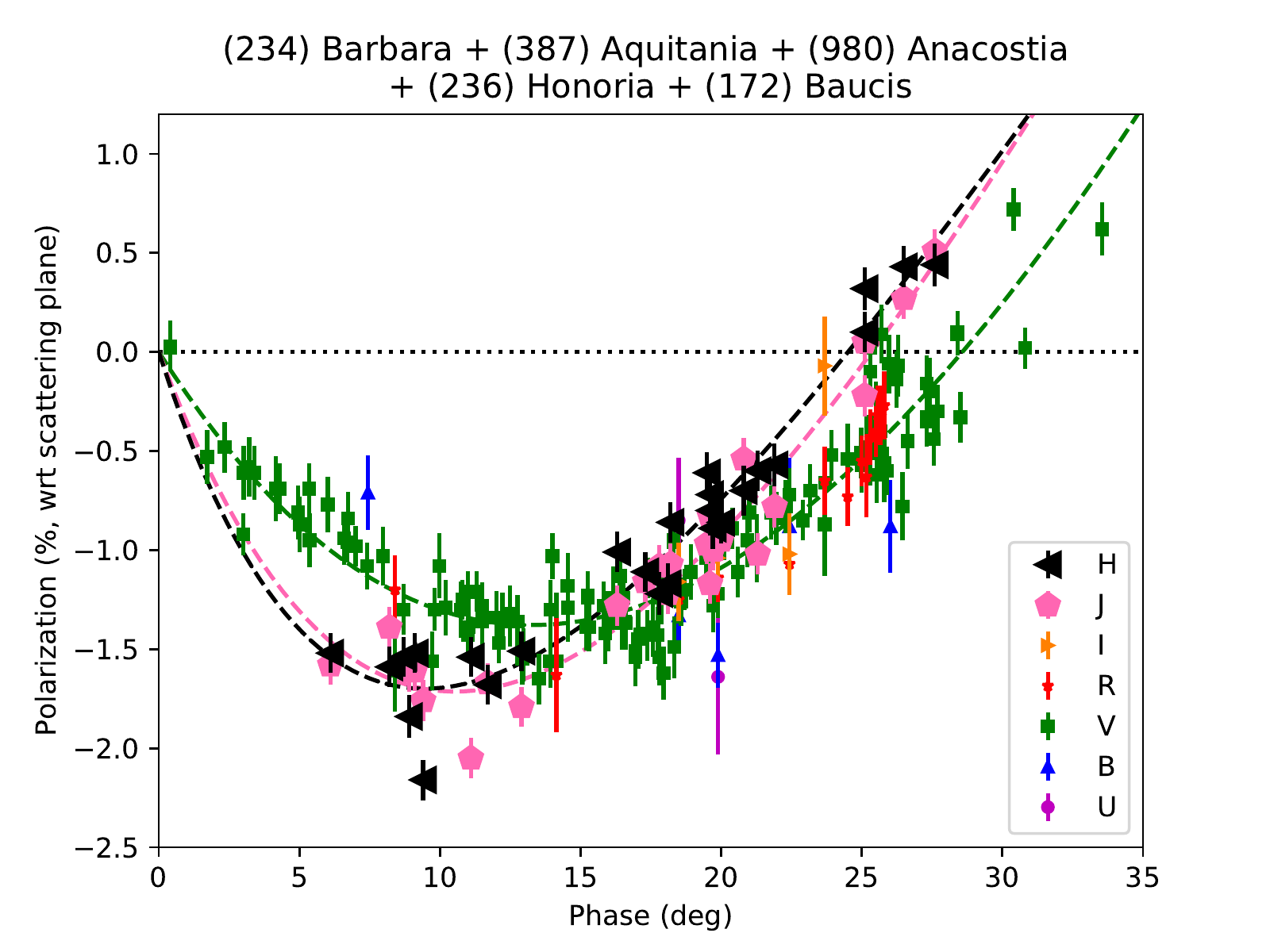}
\protect\caption{The combined polarization phase curves for five
  Barbarian asteroids. $J$ and $H$ band measurements are from WIRC+Pol and
  presented here, while other bands are drawn from the literature.
  Our observations of (402) Chloe are not included here as it follows
  a noticeably different behavior at visible wavelengths.}
\label{fig.combine}
\end{center}
\end{figure}

In Figure~\ref{fig.combine} we over-plot all data from five of our
Barbarian targets to produce a fit that is less subject to individual
measurement noise, and the best parameters of the best-fit phase
curves are presented in Table~\ref{tab.fits}.  We do not include (402)
Chloe in this plot as its $V$ band phase curve shows a different
behavior from the other Barbarians we looked at. As shown in
\citet{devogele18} Chloe has a deeper minimum polarization, a larger
P$_{min}$ angle, and larger inversion angle than the other five
objects.  \citet{bendjoya22} suggested that Barbarian objects with
deeper $P_{min}$ values may represent a distinct subclass of the
Barbarian population, something that is also suggested by the
spectroscopic classification and would result from compositional
differences between individual objects.

Even after the removal of Chloe, the V band data show a dispersion in
the polarimetric phase behavior, most noticeable in the range of phase
angles where the inversion angle is seen.  This change is also
apparent in the curve fits to the V band data shown in
Figure~\ref{fig.split}.  In contrast to this difference at $V$ band,
at $J$ and $H$ bands the polarization phase curves of all the
Barbarians we observed, including (402) Chloe, are consistent with one
another.  This suggests that the difference between these potential
subclasses may derive from compositional differences in the high index
of refraction material that dominates the visible light polarization,
but that the background matrix material that the near infrared is
probing is compositionally indistinguishable across these objects;
further investigation into the possible origin of this change is
warranted.
\clearpage

\begin{table}[ht]
\begin{center}
\scriptsize
  \noindent
  \caption{Barbarian Polarization Phase Curve Fits}
  \vspace{1ex}
  {
  \noindent
  \begin{tabular}{ccccc}
  \tableline
Band & $P_{min}$ ($\%$) & $\alpha_{min}$ (deg) & $h (\%/deg)$ & $\alpha_0$ (deg) \\
  \tableline
V & -1.38 & 13.3 & 0.17 & 28.6 \\
J & -1.71 & 10.5 & 0.19 & 25.4 \\
H & -1.70 & 9.5  & 0.18 & 24.5 \\
\hline
  \end{tabular}
  }
  \label{tab.fits}
\end{center}
  Parameters of the best-fit polarization phase curve: $P_{min}$ - the negative extrema of the curve; $\alpha_{min}$ - the phase angle where the curve reaches $P_{min}$; $h$ - the slope of the curve at the inversion angle; $\alpha_0$ - the inversion angle of the curve (where the curve crosses $P=0$).
\end{table}

\section{Discussion}

The variations we observe in the polarization phase curves for the
Barbarians point towards an interpretation that the surface elements
that dominate the polarization properties at near-infrared wavelengths
are different from the surface elements that dictate the visible-light
properties.  While the majority of polarization measurements made for
the Barbarians in the past have been at V band wavelengths, recent
data from \citet{bendjoya22} include CASLEO I band data for (234)
Barbara that is consistent with our $J$ and $H$ band measurements and
in contrast with the trend followed by the V band data.  This
points to the transition in polarization-wavelength behavior
beginning at a wavelength around $\sim700~$nm.

To interpret our observations, we can compare our findings to recent
models of dust in planetary disks. \citet{tazaki22} present results
showing that the size of the monomers that make up the aggregate dust
grains have a significant impact on the polarization behavior when the
wavelength is comparable to or smaller than the monomer size
parameter. In particular, dramatic changes occur in the polarization
behavior when $2\pi R_0 / \lambda >> 1 $ where $R_0$ is the monomer
radius and $\lambda$ is the wavelength.  Conversely when $2\pi R_0 /
\lambda << 1$ polarization is insensitive to monomer radius and shows
only small changes with wavelength.  Dramatic changes in observed
polarization, like we see here with the Barbarians, thus may be
indicative of a situation where $2\pi R_0 \sim \lambda$, revealing the
sizes of the consituant monomers that aggregated into the surface
dust. The minimal change seen in other asteroids at these wavelengths
\citep[e.g.][]{belskaya09,bendjoya22,masiero22}, would then imply that
these objects are composed of dust aggregates made of either much
smaller or much larger monomers than the Barbarians.  This could be
due to a formation effect or subsequent evolution of the body
(e.g. dust grain compaction and reduction in porosity) that was
experienced by some populations and not others.

\citet{tazaki22} also show that realistic dust monomers will have a
difference in index of refraction depending on if the included carbon
is amorphous vs organic, with amorphous carbon having a higher index
of refraction.  Additionally, while organic carbon shows nearly no
change in the real component of the index of refraction from V to H
bands, amorphous carbon's index of refraction increases with longer
wavelength.  \citet{masiero09b} investigated the effects of various
surface physical properties on the theoretical polarization phase
curve and found that while changes to the index of refraction in the
surface material produce the most dramatic effects in the location of
$\alpha_0$, changes in the size of the scattering elements produce
phase curve changes comparable to the changes we are seeing for the
Barbarians (see their Fig 2b).

The shape and size distribution of the dust particles are also
expected to have a large effect on the scattering properties of the
surface. The way that light is scattered by irregularly-shaped dust
grains is a complex combination of monomer composition and the way
that they are assembled \citep{kolokolova18}.  Wavelength has also
been shown to have a significant effect on the polarization phase
curve both in theoretical modeling \cite{min03} and for cosmic dust
analogs such as clays \citep{munoz11}.  \citet{masiero09b} considered
a single wavelength of light (600 nm) and used a simplified model of
the scattering elements that consisted of a distribution of various
sizes of spherical grains.  They found that when the scattering
elements were much larger than the wavelength the polarization curve
would have a less deep minimum, while if the minimum scattering size
was comparable to the wavelength then the depth of the $P_{min}$ and
slope at the inversion angle would be similar to what we observe for
the Barbarians in the near-infrared, as well as what we see across
wavelengths for the C-complex objects \citet{masiero22}.  While the
observed behavior of the polarization phase curve depends on the
interplay of many different physical properties of the dust, the
scattering element size approaching the observation wavelength would
thus be expected to have a large effect as the physics of the
scattering of light changes in this regime.

Given the results of these studies, we interpret our observations as
being the result of Barbarians having a two-phase surface composition:
\begin{itemize}
\item find-grained inclusions of a high albedo, high index of
  refraction material that results in visible light measurements
  showing a less deep minimum polarization and a large inversion
  angle, and
  \item a low albedo matrix material with a range of grain scales
    similar to what is observed for C-complex asteroids.
\end{itemize}
This two-phase composition would mean that the polarized photons we
receive at visible wavelengths are dominated (in terms of number) by
the fine-grained inclusions.  These inclusions become too small to
efficiently polarize light in the near-infrared, and so the polarized
photons received at those wavelengths are dominated by the matrix.

This interpretation is consistent with both the findings of
\citet{sunshine08} and \citet{devogele18}, who identify fluffy type-A
CAIs mixed with carbonaceous chondrite matrix materials as the best
fit to both the polarimetric properties and the spectra of the
Barbarians, as well as those of \citet{demeo22} who find a good
spectral match between the Barbarian (387) Aquitania and the CV3
meteorite NWA3118 \citep{relab}.  \citet{frattin22} recently have
shown that the minerals that comprise CAIs can produce Barbarian-like
visible-light phase curves in laboratory testing, further
supporting this interpretation.

Notably, as \citet{masiero09b} found, the optical size parameter that
controls the polarization is the spacing between scattering elements,
not necessarily the physical size of the individual surface grain.
Macroscopic grains with a fluffy, fractal-like composition could
result in scattering distances smaller than $\sim 1 \mu$m which would
be consistent with the observations presented here.  This also would
match previous findings on the composition of the Barbarians, as well
the connection between CV3 meteorites like NWA3118 and
Allende \citep{sunshine08}.

A similar magnitude of change in the polarization-phase curve was
produced by \citet{sultana23} for laboratory samples by changing the
ratio of olivine to FeS hyperfine grains in the sample being observed.
These observations were obtained at visible wavelengths, and olivine
was acting as a scattering element while FeS was acting as an
absorptive element in the optical mixture.  Changing from
FeS-domianted to olivine-dominated mixtures shifted the curve in a way
that is qualitatively similar to the changes we observe
(i.e. deepening P$_{min}$, lower $\alpha_0$).  This further supports a
two-phase mixture to explain the Barbarians, though shows that the
interaction between mixture, grain size, and wavelength is complex and
deserves further laboratory experimentation.

An alternate possibility for the behavior we observe could simply be
that the ratio of grain sizes of the scattering elements to wavelength
are significantly smaller in the near-infrared than in the visible.
This would be consistent with the findings of \citet{munoz21} and
specifically \citet{frattin22} who show that decreasing this ratio
results in more extreme polarization minima and maxima. An analogous
scenario was observed by \citet{hadamcik23} who measured the
polarization phase curves for materials similar to C-complex asteroids
and found that the size of the constituent grains making up the
individual scattering elements had a strong effect on the shape and
maximum of the curves.  In these cases, the observed wavelength was
constant and the grain size changed, while for our observations the
grains size distribution is constant and the wavelength is being
varied.  But in both situations, a change in the ratio between grain
size and wavelength results in a change in polarimetric properties.
However, the polarization changes produced by grain size-to-wavelength
ratio changes do not produce the shift in inversion angle we observe,
leading us to instead favor a two phase model that we describe above.
Further laboratory testing would help disciminate between these two
potential causes.

\section{Conclusions}

We present new observations of the Barbarian asteroids at
near-infrared $J$ and $H$ band wavelengths ($1.25~\mu$m and
$1.64~\mu$m respectively).  We show that the polarization-phase curves
for these objects show a dramatic change in the near-infrared
compared to the visible-light behavior, well beyond what has been seen
before.  The polariztion-phase behavior at $J$ and $H$ bands appears
to be most similar to what is observed for C-complex asteroids at
these wavelengths, despite these objects having Ld-type spectral
taxonomy most often associated with the S-complex.  Our observations
support an interpretation of the surface composition as a two-phase
mixture of CAI inclusions in a carbonaceous chondrite-like matrix.

Previous investigations have suggested that the Barbarians are a
population of very primitive survivors of an early generation of
planetesimals, based on the large fraction of CAIs on their surfaces
implied by spectra and polarimetry.  Our near-infrared data, which
reveal a dark C-complex-like matrix surrounding the CAIs, further
supports the interpretation that these are primitive objects.  While
primitive objects are often thought to have formed out beyond the ice
line in the early Solar system, where water and carbon species were
prevalent, the large compositional fraction of CAIs in the Barbarians
would instead imply formation near the hottest parts of the protosolar
disk where CAI formation occurred.  The rarity of the Barbarians may
be interpreted then as indicating they formed in a cold region of hot
inner protosolar disk, or that a transport mechanism in the forming
disk moved a large amount of CAIs from their formation zone to a
narrow region where primitive bodies were forming.  Searches for
Barbarians in populations beyond the Main Belt, as well as further
analysis of the known members of this class, will help discriminate
among these potential formation scenarios.

Although the Barbarians have shown the most dramatic changes to their
polarization at $J$ and $H$ band to date, it is likely that other
compositional classes will have unusual near-infrared polarimetric
properties.  A comprehensive survey of the variety of taxonomic types
at near-infrared wavelengths has the potential to reveal other
changing behaviors that directly probe asteroid surface mineralogy.
Polarization surveys therefore provide an important complement to
photometry and spectroscopy for understanding the composition of the
small bodies of our Solar system.

\section*{Acknowledgments}

We thank the two anonymous referees for their helpful comments that
improved this work.  JRM thanks Katherine de Kleer for helpful
discussions.  Based on observations obtained at the Hale Telescope,
Palomar Observatory as part of a continuing collaboration between the
California Institute of Technology, NASA/JPL, Yale University, and the
National Astronomical Observatories of China.  This research made use
of Photutils, an Astropy package for detection and photometry of
astronomical sources (Bradley et al.  2019).


\begin{thebibliography}{XXX}

\bibitem[Bagnulo \etal(2017)]{bagnulo17}
  Bagnulo, S., Belskaya, I., Cellino, A., \etal, 2017, European Physical Journal Plus, 132, 405.
  
\bibitem[Belskaya \etal(2009)]{belskaya09}
  Belskaya, I.N., Levasseur-Regourd, A.-C., Cellino, A., \etal, 2009, Icarus, 199, 97.

 \bibitem[Belskaya \etal(2015)]{belskaya15}
   Belskaya, I., Cellino, A., Gil-Hutton, R., \etal, 2015, Asteroids IV, ed. Michel, DeMeo, \& Bottke (Univ of Arizona Press), 151.

 \bibitem[Bendjoya \etal(2022)]{bendjoya22}
   Bendjoya, P., Cellino, A., Rivet, J.-P., \etal, 2022, A\&A, 665, A66

\bibitem[Bradley \etal(2019)]{bradley19}
  Bradley, L. \etal, 2019, astropy/photutils 0.7.2, Zenodo, doi:10.5281/zenodo.3568287

 \bibitem[Cellino \etal(2006)]{cellino06}
 Cellino, A., Belskaya, I.N., Bendjoya, Ph., \etal, 2006, Icarus, 180, 565.

\bibitem[Cellino \etal(2015)]{cellino15}
  Cellino, A., Gil-Hutton, R., \& Belskaya, I.~N., 2015, Polarimetry of Stars and Planetary Systems, 360.
 
\bibitem[DeMeo \etal(2022)]{demeo22}
  DeMeo, F.~E., Burt, B.~J., Marsset, M., \etal, 2022, Icarus, 380, 114971.

\bibitem[Devog\`{e}le \etal(2018)]{devogele18}
  Devog\`{e}le, M., Tanga, P., Cellino, A., \etal, 2018, Icarus, 304, 31.

\bibitem[Frattin \etal(2022)]{frattin22}
  Frattin, E., Martikainen, J., Mu{\~n}oz, O., \etal 2022, MNRAS, 517, 5463.

 \bibitem[Gil-Hutton \etal(2008)]{gilhutton08}
 Gil-Hutton, R., Mesa, V., Cellino, A., \etal, 2008, A\&A, 482, 309.
 
 \bibitem[Gil-Hutton \& Ca\~{n}ada-Assandri(2011)]{gilhutton11}
 Gil-Hutton, R. \& Ca\~{n}ada-Assandri, M., 2011, A\&A, 529, 86.
 
\bibitem[Hadamcik \etal(2023)]{hadamcik23}
  Hadamcik, E., Renard, J.-B., Lasue, J., \etal 2023, MNRAS, 520, 1963.

\bibitem[Hosseini(2008)]{hosseini08}
  Hosseini, S., 2008, Physica Status Solidi B Basic Research, 245, 2800.

\bibitem[Kolokolova \etal(2018)]{kolokolova18}
  Kolokolova, L., Nagdimunov, L., \& Mackowski, D.\ 2018, JQSRT, 204, 138.
  
\bibitem[Kwon \etal(2022)]{kwon22}
  Kwon, Y.~G., Masiero, J.~R., \& Markkanen, J., 2022, A\&A, 668, A97
 
\bibitem[Lupishko(2022)]{apd}  
  Lupishko, D., Ed. (2022). Asteroid Polarimetric Database V2.0. urn:nasa:pds:asteroid\_polarimetric\_database::2.0. NASA Planetary Data System
 
 \bibitem[Masiero \& Cellino(2009a)]{masiero09a}
 Masiero, J. \& Cellino, A., 2009a, Icarus, 199, 333.
 
 \bibitem[Masiero \etal(2009b)]{masiero09b}
 Masiero, J., Hartzell, C., Scheeres, D., 2009b, AJ, 138, 1557.

\bibitem[Masiero \etal(2022)]{masiero22}
  Masiero, J., Tinyanont, S., \& Millar-Blanchaer, M.A., 2022, PSJ, 3, 90.

\bibitem[Milliken(2020)]{relab}
  Milliken, R., 2020, RELAB Spectral Library Bundle V2.0.  urn:nasa:pds:relab::2.0, NASA Planetary Data System; doi:10.17189/1519032
 
 \bibitem[Millar-Blanchaer \etal(2021)]{millarblanchaer21}
   Millar-Blanchaer, M., Tinyanont, S., Jovanovic, N., \etal, 2021, SPIE, 11447, 114475Y.

 \bibitem[Min \etal(2003)]{min03}
   Min, M., Hovenier, J.~W., \& de Koter, A., 2003, JQSRT, 79-80, 939.
   
\bibitem[Muinonen \etal(2009)]{muinonen09}
Muinonen, K., Penttil\"{a}, A., Cellino, A., \etal, 2009, M\&PS, 44, 1937.

\bibitem[Mu{\~n}oz \etal(2011)]{munoz11}
  Mu{\~n}oz, O., Moreno, F., Guirado, D., \etal, 2011, Icarus, 211, 894.

\bibitem[Mu{\~n}oz et al.(2021)]{munoz21}
  Mu{\~n}oz, O., Frattin, E., Jardiel, T., \etal, 2021, ApJS, 256, 17. 

 \bibitem[Sunshine \etal(2008)]{sunshine08}
   Sunshine, J.~M., Connolly, H.~C., McCoy, T.~J., \etal, 2008, Science, 320, 514. 
   
\bibitem[Sultana \etal(2023)]{sultana23}
  Sultana, R., Poch, O., Beck, P., \etal, 2023, Icarus, in press.
  
\bibitem[Tazaki \& Dominik(2022)]{tazaki22}
  Tazaki, R. \& Dominik, C., 2022, A\&A, 663, 57.
 
\bibitem[Tinyanont \etal(2019a)]{tinyanont19a}
  Tinyanont, S., Millar-Blanchaer, M.A., Nilsson, R., \etal 2019a, PASP, 131, 25001.

\bibitem[Tinyanont \etal(2019b)]{tinyanont19b}
  Tinyanont, S., Millar-Blanchaer, M.A., Jovanovic, N. \etal 2019b, Proc. SPIE, 11132, 1113209.


\end{thebibliography}
\end{document}